\documentclass[a4paper,prb,showpacs,superscriptaddress]{revtex4}

\usepackage{epsfig}

\newcommand{\beq}{\begin{equation}}
\newcommand{\eeq}{\end{equation}}
\newcommand{\beqa}{\begin{eqnarray}}
\newcommand{\eeqa}{\end{eqnarray}}

\begin{document}


\thispagestyle{empty}
\vskip 3cm

\title{On the Structure of ${\rm ZnI_2}$.} 
\author{P. Arun}
\affiliation{Department of Physics \& Electronics, S.G.T.B. Khalsa College, 
University of Delhi, Delhi - 110 007, India}
 \email{arunp92@physics.du.ac.in}

\begin{abstract}
A new structure for ${\rm ZnI_2}$ is proposed which it exists in tetragonal
state. In this structure the ${\rm ZnI_2}$ molecule exists in a nonlinear
array and forms the basis of the tetragonal unit cell with one basis per
unit cell. The structural analysis based on the reflections listed in ASTM 
30-1479 shows that the proposed structure is correct. 
\end{abstract}

\maketitle


\section{Introduction}
The halides of cadmium, mercury and zinc form layered structures and hence
their physical and structural study is of interest for material scientist, as
is evident from the extensive studies made recently on these halides in thin 
film state by Pankaj Tyagi et al\cite{PT} and Rawat et al\cite{rawat}.
In fact ${\rm CdI_2}$ is known to crystallize in more than 200 different states 
\cite{hal}, with each poly-state exhibiting different properties. While 
${\rm CdI_2}$ in it's hexagonal state is most prevalent, ${\rm
HgI_2}$ is known to exist with a tetragonal structure. A survey of the literature
shows that ${\rm ZnI_2}$ has not attracted the same attention as ${\rm CdI_2}$ 
and ${\rm HgI_2}$. Only three structural studies have been done on ${\rm ZnI_2}$ 
\cite{Hull,french}, and they report two possible states that it can exist in. 
While Fourcroy et al\cite{french} assigns a large unit 
cell size (a=b=1.2284nm, c=2.3582nm) whose Miller indices are listed in ASTM
30-1479, the remaining studies have indexed ${\rm ZnI_2}$ with a=b=0.4388nm and 
c=1.1788nm (ASTM 10-72). Even though majority of the peak positions listed
in both the above mentioned ASTMs agree well (till the second place of
decimal), the assigned unit cell dimensions differ remarkably. This is a 
direct consequence of the fact that ASTM 10-72 only reported peaks between 
${\rm 2\theta=25^o}$ and ${\rm 80^o}$ while the more recent ASTM 30-1479 
reports X-Ray diffraction peaks between ${\rm 10^o}$ to ${\rm 60^o}$. Peaks
occurring at lower diffraction angles demand larger lattice parameters. These old 
studies have suggested ${\rm ZnI_2}$ to exist either with the same crystal 
structure as ${\rm HgI_2}$ or with a hexagonal structure similar to 
${\rm CdCl_2}$ (${\rm CdI_2}$). While ${\rm HgI_2}$ 
has a tetragonal structure with two bases per unit cell, i.e. with body center 
cubic (bcc) arrangement, ${\rm CdCl_2}$ has only one basis per it's
hexagonal unit cell. The relatively poor attention on ${\rm ZnI_2}$ and the 
fact that Fourcroy assumed a rather large basis of ${\rm Zn_4I_{10}}$ chain
motivated us to investigate whether a structure with I-Zn-I basis can 
exist.

\begin{table} 
\caption{The 'd' spacing (in \AA) of $ZnI_2$ listed in ASTM card 30-1479
along with the orignal Miller indexing (hkl) and it's error is compared with
the Miller indexing done in present work. The improvement in indexing is
evident from the decrease in error.}
\begin{center}
\begin{tabular}{ccccccccc} \hline\hline
& &\multicolumn{1}{c}{ASTM 30-1479} & & & &\multicolumn{1}{c}{Present Study} &
 &  \\
\multicolumn{1}{c}{d} & &\multicolumn{1}{c}{hkl} &  & ${\rm |error|}$ & &
\multicolumn{1}{c}{hkl} & & ${\rm |error|}$\\ \hline
6.9542 & & 112	& & 0.050154 & & 200&	& 0.005146\\
6.3406 & & 200	& & 0.180550 & & 103&	& 0.003103\\
4.5210 & & 213	& & 0.011000 & & 301&	& 0.006467\\
3.6901 & & 312	& & 0.008817 & & 133&	& 0.049893\\
3.5099 & & 224	& & 0.007706 & & 304&	& 0.010346\\
3.0797 & & 400	& & 0.000325 & & 421&	& 0.004359\\
2.9553 & & 008	& & 0.010260 & & 333&	& 0.020717\\
2.7601 & & 420	& & 0.005242 & & 431&	& 0.003713\\
2.1729 & & 440	& & 0.005006 & & 540&	& 0.002380\\
2.1267 & & 408	& & 0.001878 & & 542&	& 0.000348\\
2.0439 & & 444	& & 0.001164 & & 20(10)&& 0.001932\\
1.8442 & & 624	& & 0.005258 & & 714&	& 0.000342\\
1.7886 & & 22(12)& & 0.001330& & 33(10)&& 0.000080\\
1.7497 & & 448	& & 0.001396 & & 608&	& 0.000076\\
1.6543 & & 40(12)& & 0.001296& & 717&	& 0.001856\\
1.5358 & & 800	& & 0.004212 & & 26(10)&& 0.004041\\ 
\hline
Average Error & & & & 0.01847 & & & &0.007220 \\
\hline \hline
\end{tabular} 
\end{center}
\end{table}

\section{Results}
\par The initial indexing was done using a computer program in Turbo-BASIC 
developed by us. 
However, the refinement of cell dimensions, ${\rm \delta 2\theta}$, space group 
analysis etc., were performed using a software developed by Charles W. Burnham, 
Department of Earth and Planetary Sciences, Harvard University. The result of 
our indexing program is listed in Table I. The best indexing was obtained for 
tetragonal unit cell of dimensions a=1.3898nm and c=2.1362nm, the ratio c/a is 
equal to 1.537, and is comparable to those of layered compounds \cite{Hull}. 
These computed lattice parameters are 
comparable with the lattice parameters attributed to ${\rm ZnI_2}$ by Fourcroy, 
however are largely different from that reported in ASTM 10-72. As explained 
earlier, this is expected since ASTM 10-72 had missed out the small angle 
diffraction peaks. Table I compares the indexing of ASTM 30-1479 and the
indexing done in the present study. The error is the difference in the ASTM
listed 'd' spacing and the calculated value of 'd' using the h,k,l and unit
cell dimensions. Consistently, the error in 'd' spacing of the present study
is lower than the ASTM's reported indexing. Even though our cell dimensions are 
similar to that of Fourcroy, whose structural analysis suggested ${\rm ZnI_2}$ 
to exist with tetragonal structure having only single basis per unit cell, 
in light of the new Miller indexing we did we carried out structural analysis 
to investigate whether a structure with I-Zn-I basis can exist rather than 
${\rm Zn_4I_{10}}$ chain as described by Fourcroy.

\begin{figure}[h]
\begin{center}
\epsfig{file=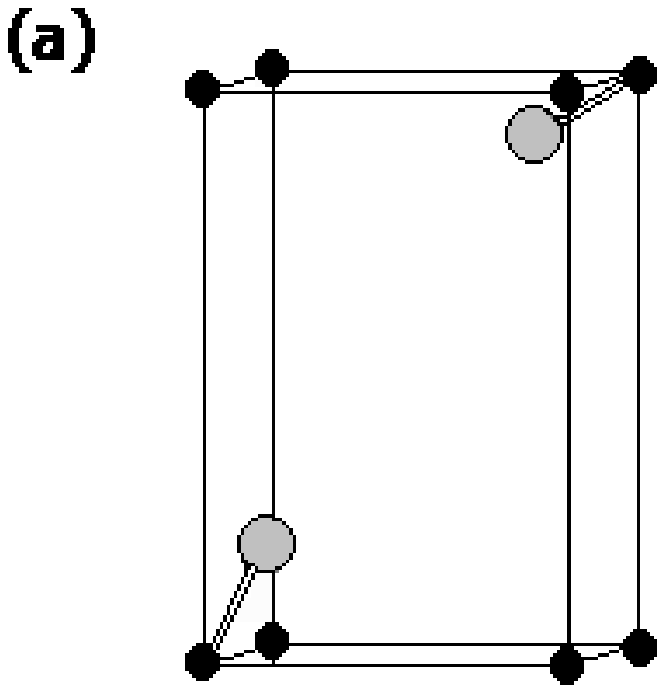,width=2.25in,height=3in}
\hfil
\epsfig{file=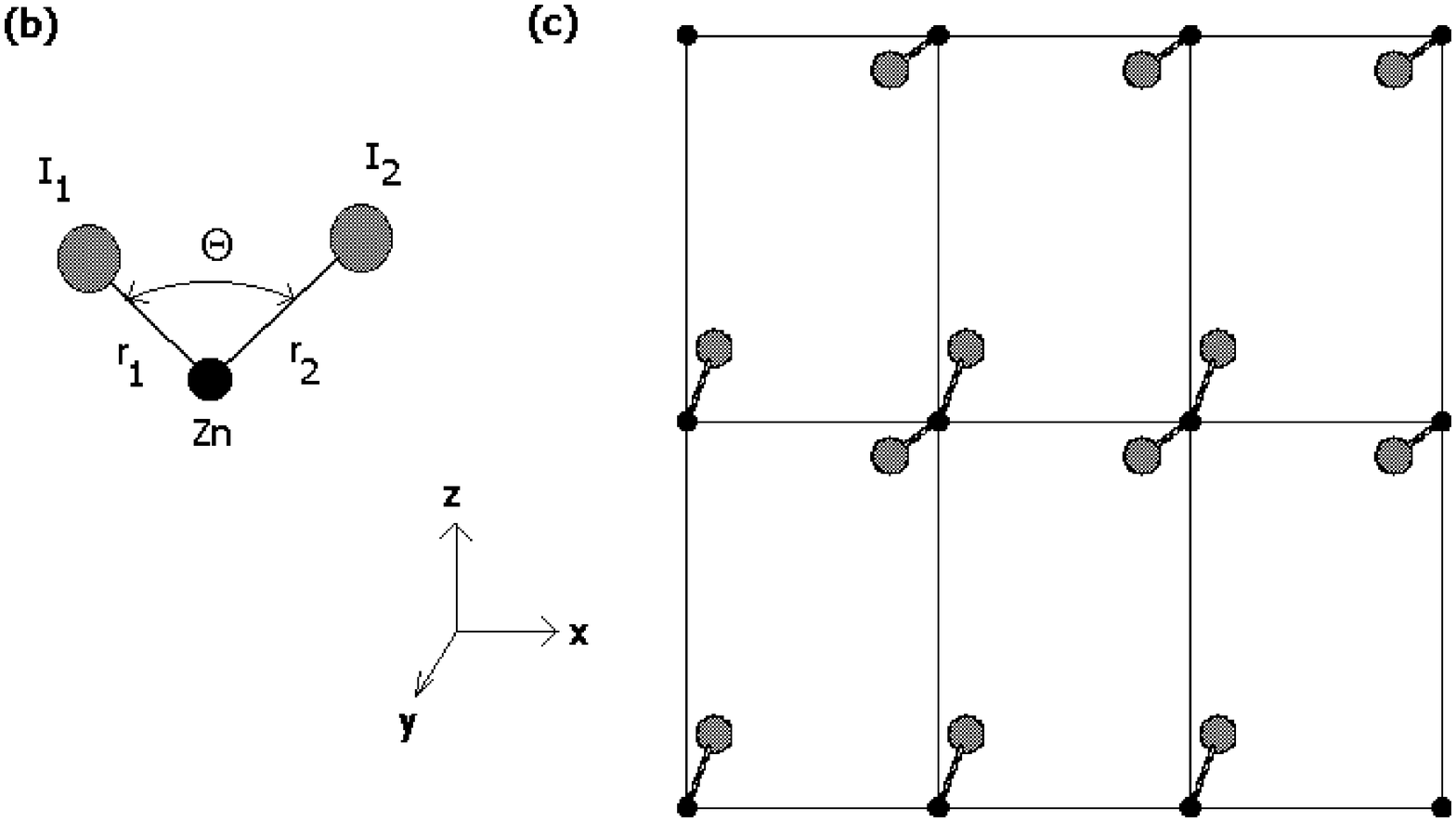, width=5.5in, height=3in}
\caption{The (a) proposed structure of ${\rm ZnI_2}$ with (b) the basis
being the non-linear molecule of ${\rm ZnI_2}$ and (c) the 'xz' plane of the
lattice formed.}
\end{center}
\label{fig:1x}
\end{figure}
\par The crystal structure determination was done using trial and error
method as outlined by Glasser in his book\cite{glass} after incoperating 
the usual corrections on the observed diffraction peak's intensity for 
polarization and Lorentz factors. Table II list the corrected intensity
(${\rm F_o}$) alongside with the measured intensity level, ${\rm I_{rel}}$. 
Figure(1a) shows the proposed 
structure. The assumed basis has one Zn atom in bonding with two I atoms 
forming a "V" shape, with the angle between the two Zn-I bonds to be ${\rm
151.1^o}$
(figure 1b). The structure gives ${\rm ZnI_2}$ a Zn-I-I-Zn-I-I-Zn 
layered structure. This is readily understood from the projections shown in 
figure(2). 

\par The initial discrepancy factor obtained
was R=0.44 or 44\%. On further refinement of the atomic positions (Iodine
atoms positions) we obtained R=0.098 or 9.8\%. This discrepancy factor may
be further reduced by including anisotropic temperature factors which
accounts for the thermal vibrations of the atoms. However, the discrepancy
is low enough to consider the proposed crystal structure to be correct. The 
atomic position of all the 
Zn and I atoms are given in Table III. The inter-atomic distances (${\rm
r_1}$ and ${\rm r_2}$ of fig) is 
easily computed from these position co-ordinates. The two bond lengths ${\rm
Zn-I_1}$ and ${\rm Zn-I_2}$ works out to be 5.827\AA\, and 7.303\AA\, 
respectively. Both bond lengths are greater then the sum of zinc and iodine 
atom's ionic radii 3.4\AA\, (1.34+2.06\AA)\cite{crc}, suggesting that the 
molecules are formed by covalent bonding. The bond lengths worked out in the
present study is far greater than those reported by Fourcroy, where the
average Zn-I bond length was only 2.6\AA. The result, hence, in that study leads 
to the conclusion that zinc and iodide atoms are held together by ionic
bonds. 

\begin{center}
{\bf Table II} Details of the diffraction data\\
\vskip 0.15cm
\begin{tabular}{cccccc} \hline\hline\\
S.No & d & ${\rm I_{rel} (\%)}$\,\, &${\rm \sqrt{I_{corr}}}$ & ${\rm |F_o|}$ & 
${\rm |F_c|}$\\
\hline\hline
1. &	6.9542\,\,\,\,	& 6&	0.96&	22.22&	23.83\\
2. &	6.3406\,\,\,\,	& 6&	0.75&	17.26&	19.51\\
3. &	4.5210\,\,\,\,	& 8&	1.23&	28.26&	27.71\\
4. &	3.6901\,\,\,\,	& 7&	1.0&	23.14&	27.86\\
5. &	3.5099\,\,\,\,	& 100&	5.68&	130.52&	122\\
6. &	3.0797\,\,\,\,	& 30&	2.53&	58.22&	66.11\\
7. &	2.9553\,\,\,\,	& 7&	1.81&	41.6&	34.61\\
8. &	2.7601\,\,\,\,	& 5&	1.16&	26.81&	10.06\\
9. &	2.1729\,\,\,\,	& 2&	1.37&	31.49&	29.74 \\
10. &	2.1267\,\,\,\,	& 45&	4.71&	108.33&	108.61\\
11. &	2.0439\,\,\,\,	& 3&	1.8&	41.45&	49.93\\
12. &	1.8442\,\,\,\,	& 10&	2.63&	60.54&	60.54\\
13. &	1.7886\,\,\,\,	& 7&	3.23&	74.36&	87.34 \\
14. &	1.7497\,\,\,\,	& 6&	3.08&	70.72&	64.55 \\
15. &	1.6543\,\,\,\,	& 3&	1.65&	37.88&	38.01\\
16. &	1.5358\,\,\,\,	& 6&	2.55&	58.72&	61.15\\ \hline \hline\\
\end{tabular}
\end{center}

\begin{center}
{\bf Table IIIa:} The position of Iodine atoms.
\vskip 0.2cm
\begin{tabular}{cccc} \hline\hline
S.No & x & y & z \\
\hline\hline
1.&	0.11\,\,\,&	0.28\,\,\,&	0.19\\
2.&	0.81\,\,\,&	0.53\,\,\,&	0.91\\
\hline\hline\\
\end{tabular} 
\end{center}
\pagebreak

\begin{center}
{\bf Table IIIb:} Position of Zinc atoms\\
\vskip 0.2cm
\begin{tabular}{cccc} \hline\hline
S.No.\,\,\, & x & y & z \\
\hline\hline
1.\,\,\,& 0 & 0 & 0 \\
2.\,\,\,& 0 & 1 & 0 \\
3.\,\,\,& 1 & 0 & 0 \\
4.\,\,\,& 1 & 1 & 0 \\
5.\,\,\,& 0 & 0 & 1 \\
6.\,\,\,& 0 & 1 & 1 \\
7.\,\,\,& 1 & 0 & 1 \\
8.\,\,\,& 1 & 1 & 1 \\
\hline\hline\\
\end{tabular} 
\end{center}

\begin{figure}[htb]
\begin{center}
\epsfig{file=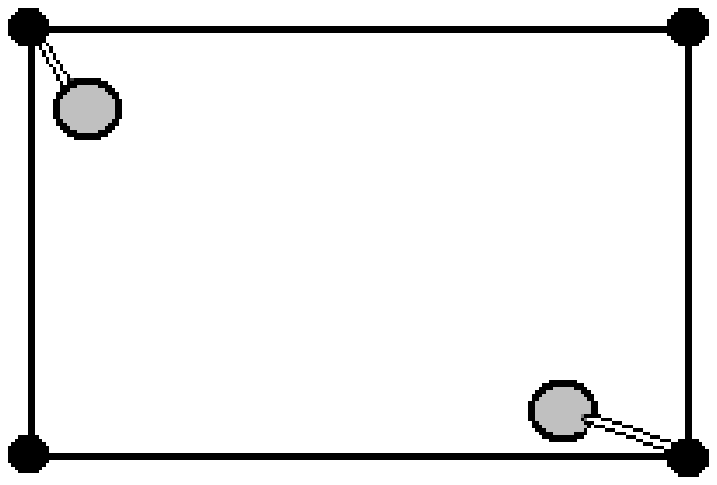,width=2in,height=1in}
\hfil
\epsfig{file=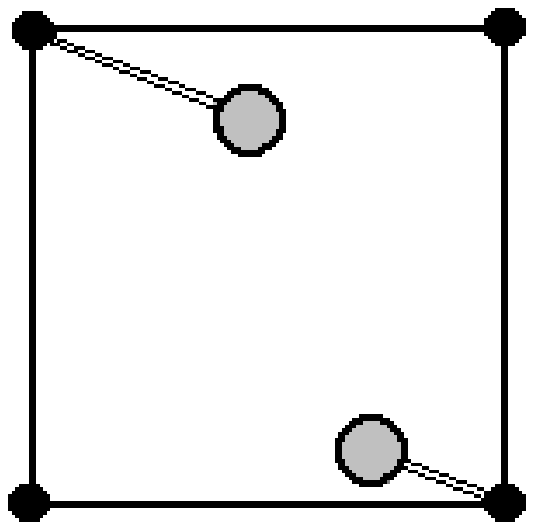,width=1in,height=1in}
\caption{Projection of ${\rm ZnI_2}$ in 'xz' (010) and 'xy' (001) plane 
respectively.}
\end{center}
\label{fig:2}
\end{figure}
\section*{Conclusion}
In conclusion, the proposed crystal structure is a possible state is which
${\rm ZnI_2}$ can exist. The large ratio of lattice parameter (c/a) along
with the larger bond lengths suggest covalent bond formation between the
atoms of the molecule. Fourcroy's structure was more of a correction
of the previous structure based on ASTM 10-72. The correction made
necesscary due to the new peaks detected at low diffraction angles. In the
present study, the indexing and lattice parmaeter's of ASTM 30-1479 have
been refined allowing for a proposing simpler structure for ${\rm ZnI_2}$
based on a simpler basis selection as compared to Fourcroy's chain of ${\rm
Zn_4I_{10}}$ basis.

\section*{Acknowledgement}
\par The help rendered by Dr. A. G. Vedeshwar, Department of Physics and
Astrophysics, Prof. P. K. Verma, Department of Geology, University of Delhi,
Delhi, India and 
Dr. R. S. Rawat, National Sciences, National Institute of Education, Nanyang
Technological University, Singapore is gratefully acknowledged. The
computational facilities extended by Mr Ashok Kumar Das is also gratefully
acknowledged.

\vfill

\end{document}